\documentstyle[12pt]{article}
\def\la{\mathrel{\mathchoice {\vcenter{\offinterlineskip\halign{\hfil
$\displaystyle##$\hfil\cr<\cr\sim\cr}}}
{\vcenter{\offinterlineskip\halign{\hfil$\textstyle##$\hfil\cr<\cr\sim\cr}}}
{\vcenter{\offinterlineskip\halign{\hfil$\scriptstyle##$\hfil\cr<\cr\sim\cr}}}
{\vcenter{\offinterlineskip\halign{\hfil$\scriptscriptstyle##$\hfil\cr<\cr\sim
\cr}}}}}
\def\ga{\mathrel{\mathchoice {\vcenter{\offinterlineskip\halign{\hfil
$\displaystyle##$\hfil\cr>\cr\sim\cr}}}
{\vcenter{\offinterlineskip\halign{\hfil$\textstyle##$\hfil\cr>\cr\sim\cr}}}
{\vcenter{\offinterlineskip\halign{\hfil$\scriptstyle##$\hfil\cr>\cr\sim\cr}}}
{\vcenter{\offinterlineskip\halign{\hfil$\scriptscriptstyle##$\hfil\cr>\cr\sim
\cr}}}}}

\begin{document}
\begin{center}
{\bf Probing Pseudo-Dirac Neutrino through Detection of Neutrino Induced Muons
from GRB Neutrinos}
\end{center}
\begin{center}
{\it Debasish Majumdar} \\
{\it Saha Institute of Nuclear Physics} \\
{\it 1/AF Bidhannagar, Kolkata 700 064, INDIA}
\end{center} 

\begin{center}
{\bf Abstract}
\end{center}

The possibility to verify the pseudo-Dirac nature of neutrinos is 
investigated here via the detection of ultra high energy neutrinos 
from distant cosmological objects like GRBs. The very long baseline 
and the energy range from $\sim$ TeV to $\sim$ EeV for such neutrinos
invokes the likelihood to probe very small pseude-Dirac splittings. 
The expected secondary muons from such neutrinos that can be detected 
by a kilometer scale detector such as ICECUBE is calculated. The pseudo-Dirac
nature, if exists, will show a considerable departure from flavour oscillation
scenario in the total yield of the secondary muons induced by such 
neutrinos.

\newpage

\section{Introduction}
Evidence has been obtained from the satellite-borne observations, the existence
of the Gamma Ray Bursts (GRB) from extra galactic (or galactic) sources
characterised by sudden intense flashes of $\gamma$-rays. The GRBs can produce
very high energy ($\ga 10^{16}$ eV) cosmic rays. Although the mechanism
of such GRBs are not fully understood but various model calculations
suggest that neutrinos of that energy range are also produced in GRBs and
should be detected by very large detectors like a kilometer cube water
cherenkov detector (e.g. ICECUBE at south pole). Because of their
astronomically long
baseline ($\ga$ Mega Parsec) these ultra high energy (UHE) neutrinos
may open up a new window in very small mass square difference ($\Delta m^2$)
regime and may throw more insight to neutrino physics. A viable way to detect
such UHE neutrinos is to look for upward going secondary muons produced by
the charged current (CC) interaction of such UHE $\nu_\mu$ with the 
terrestrial rock. Already installed AMANDA in the
south pole and the future ICECUBE \cite{icecube} detector,  
are able to detect such
secondary muons. Both are water Cerenkov detectors and use south pole
ice as the detector material.

The possibility that the UHE neutrinos from a distant GRB can probe very 
low $\Delta m^2$ region much below
the solar or atmospheric neutrino regime, can be utilised in investigating 
the proposed pseudo-Dirac nature of the neutrinos \cite{wolf,kob} through 
its detection in large ($\sim$ Km$^2$) detectors \cite{beacom} like ICECUBE.  
In the 
pseudo-Dirac scenario, each active neutrino is split into two almost 
degenerate components
of active and sterile part. This can be theoretically realised from the
generic mass matrix in [$\nu_L, (\nu_R)^C$] basis which can be written
as (for one generation)
$$
\left ( \begin{array}{cc} m_L & m_D \\ m_D & m_R \end{array} \right ),
$$
where the tiny Majorana mass terms $m_L$ and $m_R$ ($m_L, m_R << m_D$)
are introduced to slightly lift the degeneracy of Dirac mass $m_D$. Thus
the Dirac neutrino is split into a pair of almost degenerate Majorana
neutrinos with nearly maximum mixing angle given by 
$\tan 2\theta = 2m_D/(m_R - m_L)$. In this scenario, therefore, 
each of the three neutrino species is split up into a pair of
neutrinos with  very tiny mass square difference 
$\Delta m^2 = 2m_D(m_L+m_R)$. For three generation scenario, 
therefore each of the three types 
of neutrinos has this pseudo-Dirac splitting of very small mass 
square difference.

The purpose of this work is to demonstrate the possibility of 
probing the pseudo-Dirac nature by studying the neutrino 
oscillation effects through the detection yield of neutrino induced 
muons in a large (Km$^2$) neutrino detector such as ICECUBE. 
For this purpose, the expected muon signal induced by neutrinos 
from a GRB at Mpc distance is calculated 
by separately folding (i) the flavour oscillation effects 
and (ii) the oscillation effects in pseudo-Dirac scenario to 
the GRB neutrino flux. 

\section{GRB flux, neutrino oscillations and number of secondary muons}

The GRB neutrino flux is estimated considering the 
the relativistic fireball model \cite{waxman}. 
In the relativistic expanding fireball model of GRB, protons (also electrons,
positrons and photons) produced 
in the magnetic field of the rotating accretion disc around a possible
black hole are accelerated perpendicular
to the accretion disc at almost the speed of light forming a 
jet which is referred to as fireball. The burst is supposed to be
the dissipation of kinetic energy of this relativistic expanding 
fireball. 
These very highly energetic protons in the jet then interact with the 
photons and produce pions (cosmological beam dump) through the process
of $\Delta$ resonance. (The pions are also produced through the $pp$ 
process). These pions then decay to yield $\nu_\mu$ and 
$\nu_e$ in the approximate proportion of 2:1. 

The neutrino flux from a GRB depends on several GRB parameters. 
Firstly, the Lorentz boost factor $\Gamma$ is required for the transformation 
from the fireball blob to observer's frame of reference. As the 
shocked protons in the blob photoproduce pions, the photon break 
energy (as the photon spectrum is considered broken) and the photon luminosity
$L_\gamma$ (generally $\sim 10^{53}$ ergs/sec) are important parameters 
for determining the neutrino spectrum. 

With all these, the neutrino spectrum from a GRB can be parametrised 
as \cite{nayan,pijush}
\begin{equation}
\frac {dN_\nu} {dE_\nu} = A \times {\rm min}(1,E_\nu/E_\nu^b) \times 
\frac{1} {E_\nu^2}.
\end{equation}
In the above, $E_\nu$ is the neutrino energy $N_\nu$ is the number of 
neutrinos and
\begin{eqnarray}
E_\nu^b &\simeq& 10^6 \frac {\Gamma_{2.5}^2} 
{E_{\gamma,{\rm MeV}}^b} {\rm GeV} \nonumber \\
\Gamma_{2.5} &=& \Gamma/10^{2.5} \nonumber \\
A &=& \frac {E_{\rm GRB}} {1 + \ln(E_{\nu{\rm max}}/E_\nu^b)}\,\,,
\end{eqnarray}  
where $E_{\nu{\rm max}}$ is the cut-off energy for the GRB neutrinos and 
$E_{\rm GRB}$ is the total energy that a GRB emits. Now, the observed 
energy $E_\nu^{\rm obs}$ of a neutrino with the actual energy $E_\nu$ 
coming from a GRB at a redshift distance $z$ is given by the relation 
$E_\nu^{\rm obs} = E_\nu/(1+z)$ and similarly, the maximum observable 
neutrino energy $E_{\nu{\rm max}}^{\rm obs}$ is 
$E_{\nu{\rm max}}^{\rm obs} = E_{\nu{\rm max}}/(1+z)$. The comoving 
distance $d$ of a GRB at redshift $z$ is given by 
\begin{equation}
d(z) = \frac {c} {H} \int_0^z \frac {dz^\prime} 
{\sqrt{\Omega_\Lambda + \Omega_M((1+z^\prime)^3}}
\end{equation}  
where $\Omega_M$ is the matter density (both luminous and dark), 
$\Omega_\Lambda$ is the dark 
energy density respectively in units of critical density of 
the universe and c,H 
are the velocity of light in vacuum and Hubble constant respectively.
In the present calculation $c = 3 \times 10^5$ Km/sec and 
$H = 72$ Km/sec/Mpc (1 Mpc $= 3.086 \times 10^{19}$ Km). Therefore the 
neutrinos from a single GRB that can be observed on earth per unit 
energy per unit area of the earth is given by,
\begin{equation}
\frac {dN_\nu^{\rm obs}} {dE_\nu^{\rm obs}} = 
\frac {dN_\nu} {dE_\nu} \frac {1} {4\pi d^2(z)} (1+z)
\end{equation}  
Here, as is evident from the above, the total flux per unit area per 
unit energy from a single GRB is being considered rather than the flux  
per unit time. 
  
The production process of UHE neutrinos suggests that the neutrino
flavours are produced in the ratio $\nu_e : \nu_\mu : \nu_\tau = 
1:2:0$. Considering the maximal mixing between $\nu_\mu$ and $\nu_\tau$
(as indicated by the atmospheric neutrino data) and the element 
$U_{e3}$ of the mass-flavour mixing matrix to be zero, the flavour 
ratio on reaching the earth, for neutrino mass-flavour oscillation, becomes
$\nu_e : \nu_\mu : \nu_\tau = 1:1:1$ irrespective of the solar mixing
angle. Needless to say, because of the astronomical baseline 
($L \sim {\rm Mpc}$), 
the acquired relative phases of the propagating neutrino mass eigenstates
is averaged out ($\Delta m^2L/E >> 1$) and the UHE neutrinos from a 
GRB reaching the earth are incoherent mixture of mass eigenstates. 
Therefore, the probability for measuring a particular flavour 
$\beta$ by a terrestrial neutrino telescope, if only flavour 
oscillation is considered, is $P_\beta = 1/3$.

On the other hand, in pseudo-Dirac scenario, we have each of the three 
mass eigenstates $\nu_1$, $\nu_2$ and $\nu_3$ to be nearly degenerate 
pairs and thus one obtains a total of six mass eigenstates. Kobayashi and 
Lim \cite{kob} worked out the mixing in such scenario and calculated 
the oscillation probability. Follwing \cite{kob} and Beacom 
\cite{beacom} the probability $P_\beta$ to detect a flavour 
$\beta$ by a neutrino
telescope for pseudo-Dirac neutrinos is 
\begin{equation}
P_\beta = \displaystyle\sum_\alpha w_\alpha \sum_{j=1}^3 
|U_{\alpha j}|^2 |U_{\beta j}|^2 \left [ 1 - {\rm sin}^2
\left (\frac {\Delta m_j^2 L} {4 E} \right ) \right ]\,\, ,
\end{equation}
where $m_j (j=1,3)$ denotes the mass eigenstates for the three
types of neutrinos, $\Delta m_j^2$ is the mass square difference
due to pseudo-Dirac splitting of the mass eigenstate $\nu_j$. 
$\alpha$, $\beta$, .. denote the flavour index and $U_{\alpha j}$ 
is the CKM matrix for three generation mass to flavour mixing.
$w_\alpha$ is the relative flux of 
each of the neutrino flavours ($\alpha$) at the production point
($\sum_\alpha w_\alpha = 1$).     

The total number of secondary muons induced by GRB neutrinos at a
detector of unit area is given by (following \cite{gaisser,raj1,nayan})
\begin{equation}
S = \int_{E_{\rm thr}}^{E_{\nu{\rm max}}^{\rm obs}} 
dE_\nu^{\rm obs} \frac {dN_{\nu}^{\rm obs}} {dE_\nu^{\rm obs}} P_{\rm surv}
(E_\nu^{\rm obs},\theta_z) P_\mu(E_\nu^{\rm obs},E_{\rm thr})
\end{equation}
In the above, $P_{\rm surv}$ is the probability that a neutrino reaches 
the detector without being absorbed by the earth. This is a function of 
the neutrino-nucleon interaction length in the earth and the effective 
path length $X(\theta_z)$ (gm cm$^{-2}$) for incident neutrino 
zenith angle $\theta_z$
($\theta_z = 0$ for vertically downward entry with respect to the detector).
The interaction length $L_{\rm int}$ is given by 
\begin{equation}
L_{\rm int} =  \frac {1} {\sigma^{\rm tot}(E_\nu^{\rm obs}) N_A}
\end{equation}
and 
\begin{equation}
P_{\rm surv} (E_\nu^{\rm obs},\theta_z) = \exp [-X(\theta_z)/L_{\rm int}]  
= \exp [-X(\theta_z) \sigma^{\rm tot} N_A ].
\end{equation}
where $N_A (= 6.022 \times 10^{23} {\rm gm}^{-1})$ is the Avogadro number
and $\sigma^{\rm tot} (= \sigma^{\rm CC} + \sigma^{\rm NC})$ is the 
total cross section. The effective path length $X(\theta_z)$ is 
calculated as 
\begin{equation}
X(\theta_z) = \int \rho (r(\theta_z, \ell) d\ell.
\end{equation}
In Eq. (9), $\rho (r(\theta_z, \ell)$ is the matter density inside the earth
at a distance
$r$ from the centre of the earth for neutrino path length $\ell$ entering
into the earth with a zenith angle $\theta_z$.
The quantity $P_\mu(E_\nu^{\rm obs},E_{\rm thr})$ in Eq. (6) is the 
probability that a secondary
muon is produced by CC interaction of $\nu_\mu$ and reach the 
detector above the threshold energy $E_{\rm thr}$. This is then a 
function of $\nu_\mu N$ (N represents nucleon) - CC interaction 
cross section $\sigma^{\rm CC}$
and the range of the muon inside the rock. 
\begin{equation}
P_\mu(E_\nu^{\rm obs},E_{\rm thr}) = N_A \sigma^{\rm CC} 
\langle R(E_\nu^{\rm obs};E_{\rm thr})
\rangle
\end{equation}
In the above $\langle R(E_\nu^{\rm obs};E_{\rm thr})\rangle$ 
is the average muon range given by
\begin{equation}
\langle R(E_\nu^{\rm obs};E_{\rm thr}) \rangle = \frac {1} {\sigma^{\rm CC}}
\displaystyle\int_0^{1 - E_{\rm thr}/E_\nu} 
dy R(E_\nu^{\rm obs} (1 - y), E_{\rm thr})
\frac {d\sigma^{\rm CC}(E_\nu^{\rm obs},y)} {dy}
\end{equation}
where $y = (E_\nu^{\rm obs} - E_\mu)/E_\nu^{\rm obs}$ is the 
fraction of energy loss 
by a neutrino of energy $E_\nu^{\rm obs}$ in the charged current production of
a secondary muon of energy $E_\mu$. Needless to say that a muon thus produced
from a neutrino with energy $E_\nu$ can have the 
detectable energy range between
$E_{\rm thr}$ and $E_\nu$. The range $R (E_\mu, E_{\rm thr})$ for a muon
of energy $E_\mu$ is given as
\begin{equation}
R (E_\mu, E_{\rm thr}) = \displaystyle\int^{E_\mu}_{E_{\rm thr}} 
\frac {dE_\mu} {\langle dE_\mu/dX \rangle} \simeq \frac {1} {\beta}
\ln \left ( \frac {\alpha + \beta E_\mu} {\alpha + \beta E_{\rm thr}}.
\right )
\end{equation}
The average lepton energy loss with energy $E_\mu$ per unit distance 
travelled is given by 
\cite{gaisser} 
\begin{equation}
\left \langle \frac {dE_\mu} {dX} \right\rangle = -\alpha - \beta E_\mu
\end{equation}
The values of $\alpha$ and $\beta$ used in the present calculations
are 
\begin{eqnarray}
\alpha &=& \{ 2.033 + 0.077\ln[E_\mu {\rm (GeV)}] \}\times 10^{-3} {\rm GeV}
{\rm cm}^2 {\rm gm}^{-1} \nonumber \\
\beta &=& \{ 2.033 + 0.077\ln[E_\mu {\rm (GeV)}] \} \times 10^{-6}
{\rm cm}^2 {\rm gm}^{-1} 
\end{eqnarray}
for $E_\mu \la 10^6$ GeV \cite{dar} and 
\begin{eqnarray}
\alpha &=& 2.033 \times 10^{-3} {\rm GeV} 
{\rm cm}^2 {\rm gm}^{-1} \nonumber \\
\beta &=& 3.9 \times 10^{-6} 
{\rm cm}^2 {\rm gm}^{-1}
\end{eqnarray}
otherwise \cite{guetta1}.  

\section{Calculations and results}
The GRB neutrino flux is calculated for a GRB with energy 
$E_{\rm GRB} = 10^{53}$ ergs. The neutrino break energy $E_\nu^b$ is 
calculated following Eq. (2) with the Lorentz factor $\Gamma = 50.12$ 
and corresponding photon break energy $E_{\gamma,{\rm MeV}}^b = 0.794$. 
These values are obtained from Guetta et al \cite{guetta2} from their
fireball model framework calculations. These values are tabulated 
in Ref. \cite{nayan}. The calculation is performed for several 
values of redshift ($z$). In this calculation all length units 
are taken in Km.

The probability $P_\beta$ in Eq. (5) is computed with solar mixing 
angle $\theta_\odot = 32.31^o$, the atmospheric mixing angle 
$\theta_{\rm atm} = 45.0^o$ and 1-3 mixing angle $\theta_{13} = 0$.
Also, the representative pseudo-Dirac splittings are considerd as 
$\Delta m_j^2 = 10^{-12}$ eV$^2$ for each of the three species.
Note that, in this case, 
$P_\beta$ for mass-flavour oscillation is 1/3.
GRB neutrino flux with redshift $z$, per square kilometer on 
earth per unit energy (GeV) for a zenith angle $\theta_z$, 
is obtained using Eq. (1 -4) and 
then multiplied by the probability $P_\beta$. 

The secondary muon yield at a kilometre scale detector such as 
ICECUBE is calculated using Eqs. (6 - 15). The earth matter density 
in Eq. (9) is taken from \cite{raj1} following the Preliminary 
Earth Reference Model (PREM). The interaction cross-sections - both charged 
current and total - used in these equations are taken from the 
tabulated values (and the analytical form) given in Ref. \cite{raj2}.
In the present calculations $E_{\nu{\rm max}}^{\rm obs} = 10^{11}$ GeV and 
threshold energy $E_{\rm thr} = 1$ TeV are considered.   

Fig. 1 shows the total yield of secondary muons induced 
by UHE neutrinos from GRBs of different red shift values ranging
from 0.02 to 3.8, for three
cases namely (i) no neutrino oscillation (ii) mass-flavour oscillation
and (iii) the psedo-Dirac oscillation. The zenith angle is taken to be 
$\theta_z = 180^o$ (vertically upwards neutrino, i.e. muons coming 
from vertically below the detector). Fig. 2 is same as Fig. 1 but for 
$\theta_z = 100.9^o$. The pseudo-Dirac case distinctly differs from 
flavour oscillation and no oscillation scenario. 
In the present case considered here, the secondary muon events 
for pseudo-Dirac oscillation is  
almost half the yield from expected events if
the UHE neutrinos suffers only mass-flavour oscillation and 
same is around six times less in case there is no oscillation.
In Fig. 3, the secondary muon events in each bin of $E_\nu^{\rm obs}$ are
shown. The events below 10$^{-15}$ are not shown in Fig. 3 for clarity.

\section{conclusion}
In this report, the possibility of probing pseudo-Dirac neutrino 
through ultra high energy neutrinos from distant GRBs is considered.
It is demonstrated, that such UHE neutrinos from GRB are indeed capable
of producing oscillation effects clearly distinct from flavour oscillation
or no oscillation of such neutrinos. This can be realised by looking
at the yield of secondary muons, produced by these neutrinos, 
at a kilometer scale detector such as ICECUBE at 
south pole. Detecting UHE neutrinos is also a viable means to probe
as also verify the pseudo-Dirac splitting if it really exists.

\newpage
\begin{center}
{\bf Figure Captions}
\end{center}

\noindent Fig. 1 The comparison of total secondary muon yields in cases of 
(i) pseudo-Dirac oscillation (ii) Mass-Flavour oscillation and (iii) 
No oscillation for ultra high energy neutrinos from GRBs at different
zshift $(z)$ distance. The zenith angle $\theta_z$ is 180$^o$. See text
for other GRB parameters considered.

\noindent Fig. 2 Same as Fig. 1 but for $\theta_z = 100.9^o$.

\noindent Fig. 3 The secondary muon events in each bin of $E_\nu^{\rm obs}$.      
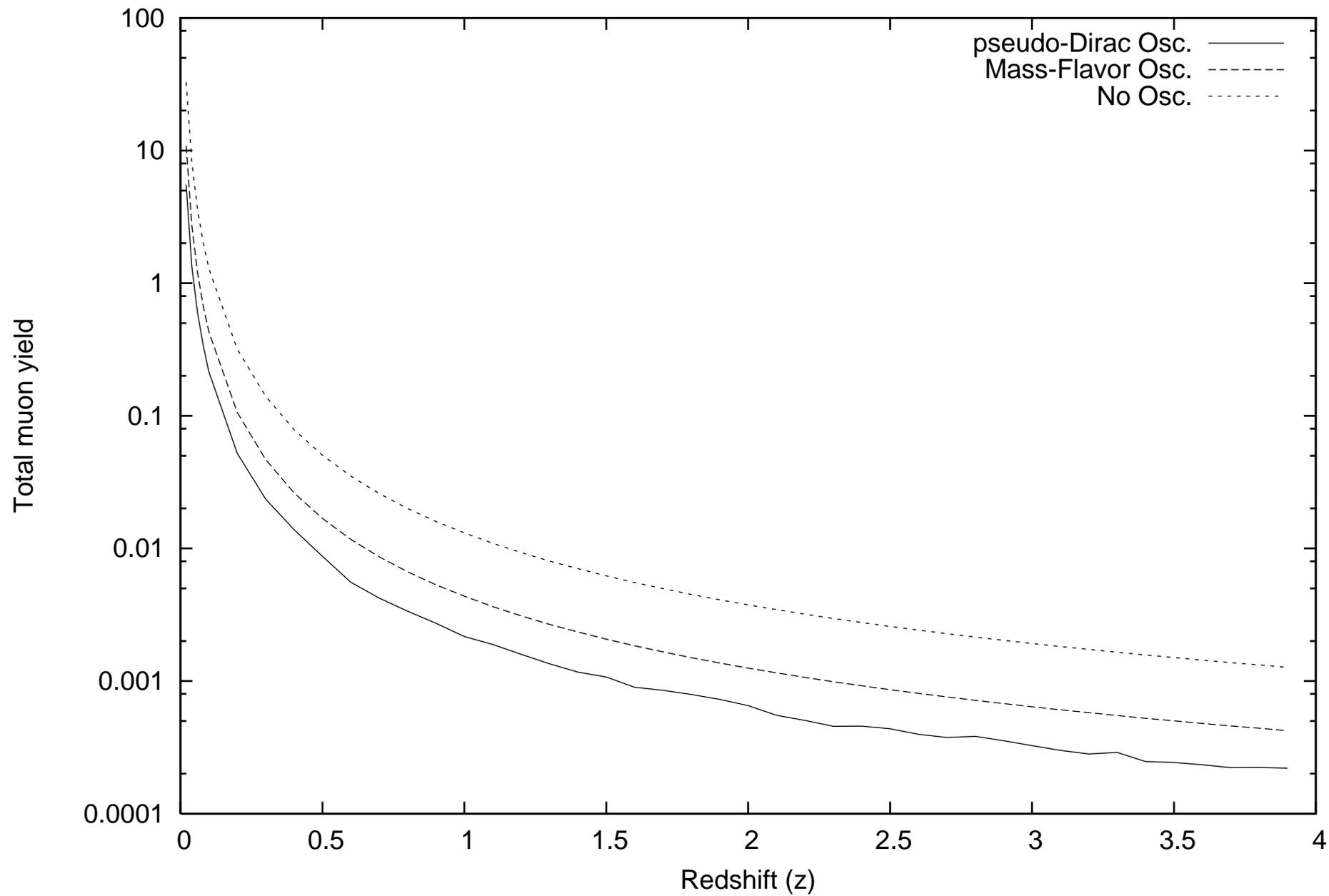
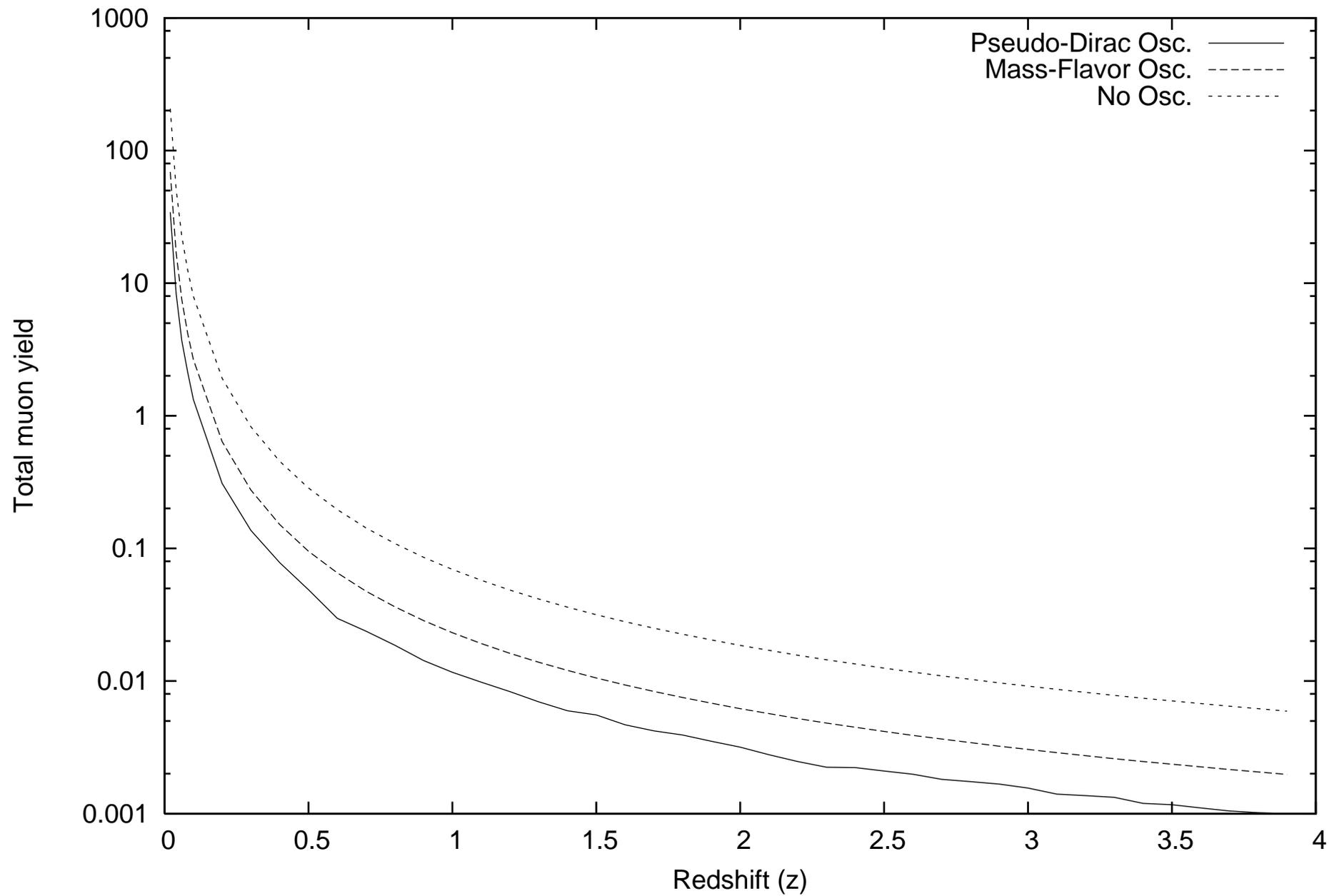
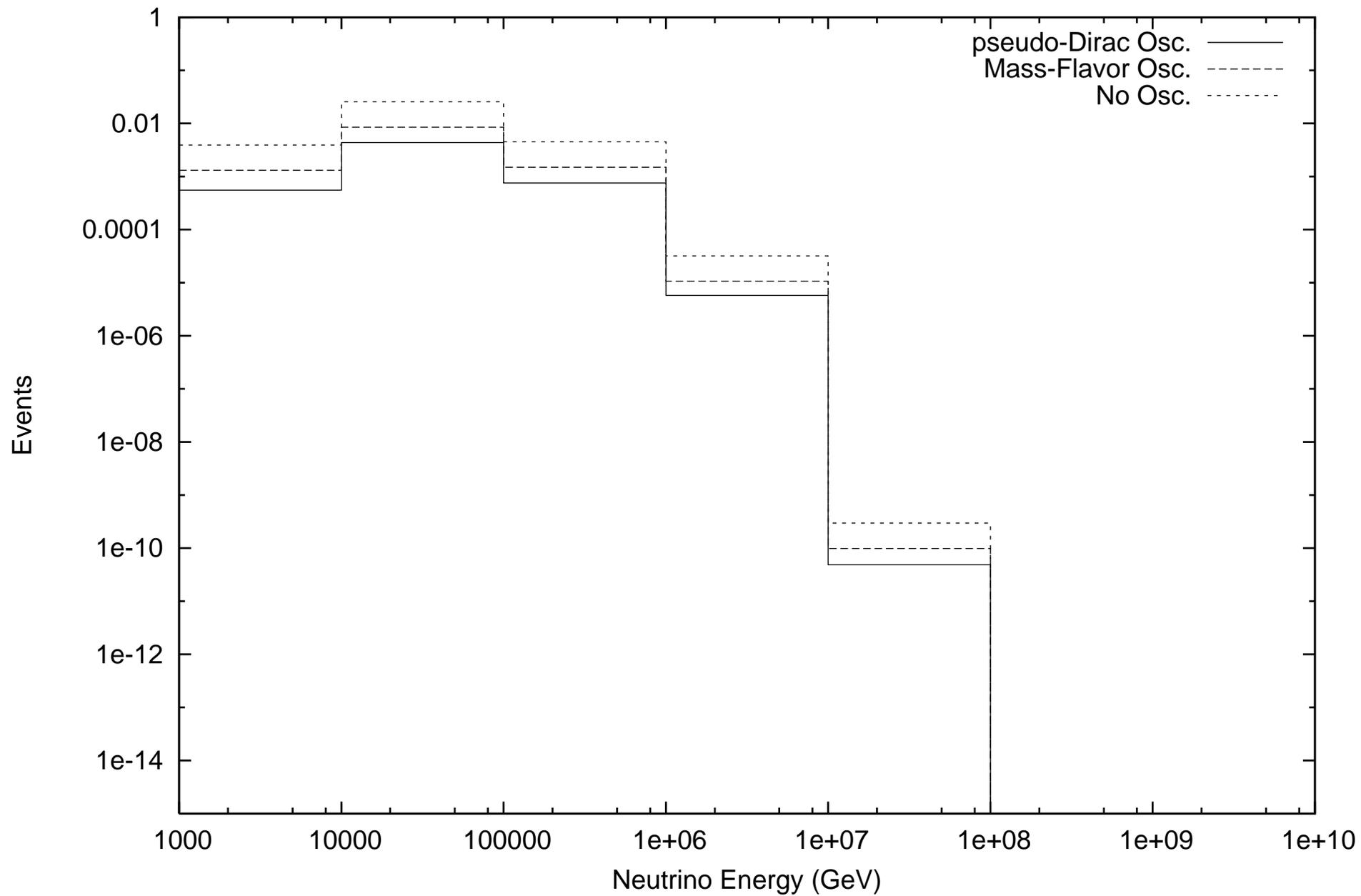
\end{document}